\documentclass[journal=nalefd,manuscript=article]{achemso}

\usepackage{amsmath, amsthm, amssymb}
\usepackage{graphicx}
\usepackage{xcolor}
\usepackage{xargs}
\usepackage[utf8]{inputenc}
\usepackage{ragged2e}

\usepackage{cleveref}
\crefname{equation}{Eq.}{Eqs.}
\usepackage{natbib}
\usepackage{soul}
\usepackage{subfigure}
\usepackage{graphicx}
 
\author{Yu-Jie Zhong}
\affiliation{Department of Physics, National Cheng Kung University, Tainan 70101, Taiwan}
\alsoaffiliation{Center for Quantum Frontiers of Research and Technology (QFort), National Cheng Kung University, Tainan 70101, Taiwan}

\author{ Jia-Cheng Li }
\affiliation{Program on Key Materials, Academy of Innovative Semiconductor and Sustainable Manufacturing, National Cheng Kung University, Tainan 70101, Taiwan}
\alsoaffiliation{Center for Quantum Frontiers of Research and Technology (QFort), National Cheng Kung University, Tainan 70101, Taiwan}

\author{Xuan-Fu Huang}
\affiliation{Department of Physics, National Cheng Kung University, Tainan 70101, Taiwan}
\alsoaffiliation{Center for Quantum Frontiers of Research and Technology (QFort), National Cheng Kung University, Tainan 70101, Taiwan}

\author{Ying-Je Lee}
\affiliation{Department of Physics, National Cheng Kung University, Tainan 70101, Taiwan}
\alsoaffiliation{Center for Quantum Frontiers of Research and Technology (QFort), National Cheng Kung University, Tainan 70101, Taiwan}

\author{Ting-Zhen Chen}
\affiliation{Department of Physics, National Cheng Kung University, Tainan 70101, Taiwan}
\alsoaffiliation{Center for Quantum Frontiers of Research and Technology (QFort), National Cheng Kung University, Tainan 70101, Taiwan}

\author{ Jia-Ren Zhang}
\affiliation{Department of Physics, National Cheng Kung University, Tainan 70101, Taiwan}
\alsoaffiliation{Center for Quantum Frontiers of Research and Technology (QFort), National Cheng Kung University, Tainan 70101, Taiwan}

\author{Angus Huang}
\affiliation{Department of Physics, National Tsing Hua University, Hsinchu 30013, Taiwan}

\author{Hsiu-Chuan Hsu}
\affiliation{Graduate Institute of Applied Physics, National Chengchi University, Taipei 11605, Taiwan}
\email{hcjhsu@nccu.edu.tw}

\author{Carmine Ortix}
\affiliation{Dipartimento di Fisica ``E. R. Caianiello'', Universit\`a di Salerno I-84084 Fisciano (Salerno), Italy}

\author{Ching-Hao Chang}
\affiliation{Department of Physics, National Cheng Kung University, Tainan 70101, Taiwan}
\alsoaffiliation{Center for Quantum Frontiers of Research and Technology (QFort), National Cheng Kung University, Tainan 70101, Taiwan}
\alsoaffiliation{Program on Key Materials, Academy of Innovative Semiconductor and Sustainable Manufacturing, National Cheng Kung University, Tainan 70101, Taiwan}
\email{cutygo@phys.ncku.edu.tw}

\title{Large positive magnetoconductance in carbon nanoscrolls}


\begin{document}

\begin{abstract}
We theoretically demonstrate that carbon nanoscrolls -- spirally wrapped graphene layers with open endpoints -- can be characterized by a large positive magnetoconductance. We show that when a carbon nanoscroll is subject to an axial magnetic field of  several Tesla, the ballistic conductance at low carrier densities of the nanoscroll has an increase of about 200\%. Importantly, we find that this positive magnetoconductance is not only preserved in an imperfect nanoscroll (with disorder or mild inter-turn misalignment) but can even be enhanced in the presence of on-site disorder.  We prove that the positive magnetoconductance comes about the emergence of magnetic field-induced zero energy modes, specific of rolled-up geometries. Our results establish curved graphene systems as a new material platform displaying  sizable  magnetoresistive phenomena. 

\begin{figure}[htb]
    \includegraphics[width=0.7\textwidth]{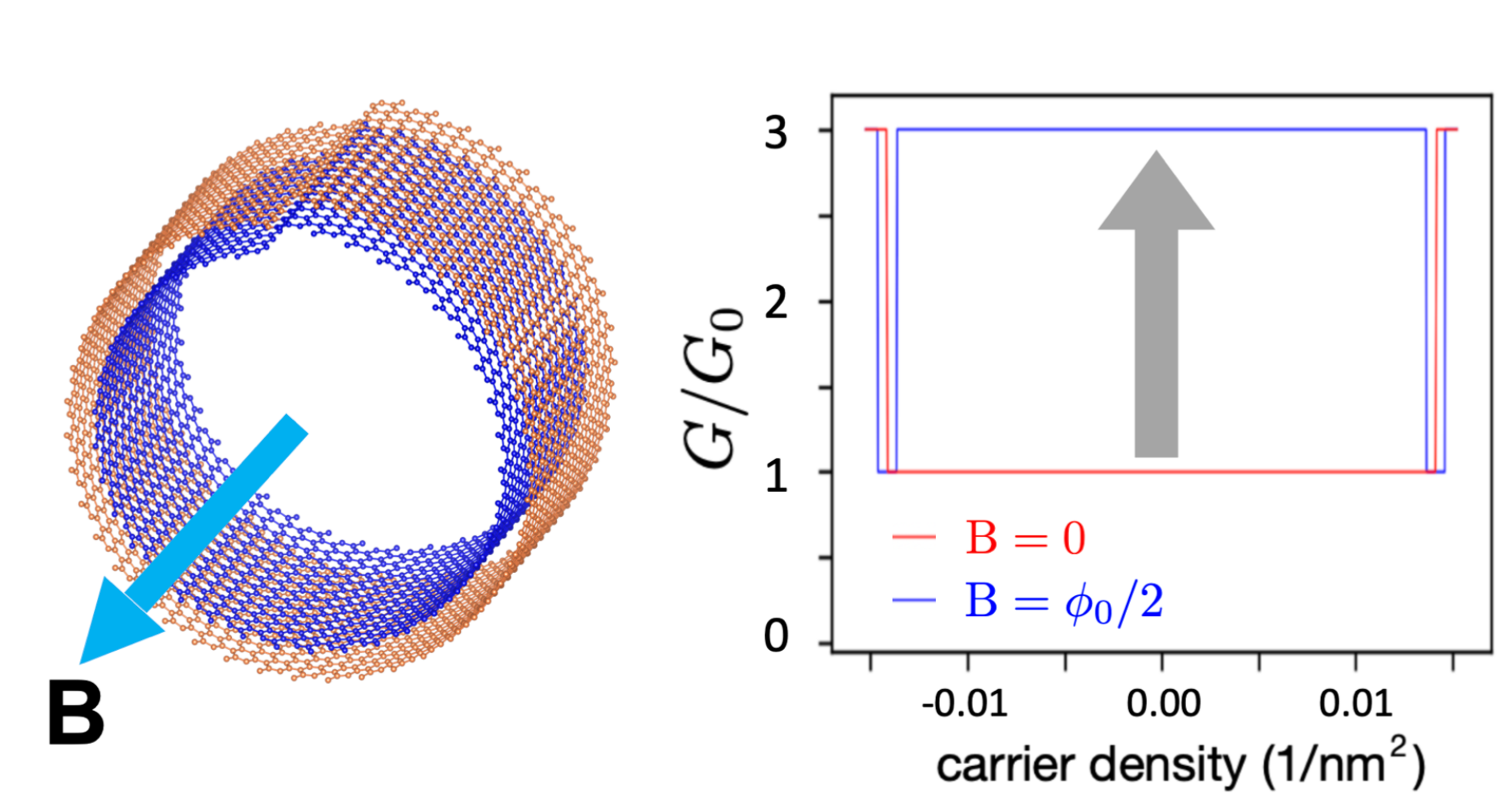}
    \centering
    \label{Abstractfig}
    \end{figure}
\end{abstract}

\maketitle

\section{Introduction}
Magnetocondutance -- the change of conductance in response to an externally applied magnetic field -- appears in {different} magnetic and non-magnetic materials alike and can have various physical origins. At very low temperatures, the presence of quantum interference effects, specifically weak (anti)localization, leads to a positive (negative) magnetoconductivity~\cite{Rammer1991}. Weak antilocalization has been recently observed for instance in topological insulators~\cite{Lu2014} and is related to the strongly spin-orbit coupled Dirac surface states of these materials. 
The competition between weak localization and antilocalization in InGaAs-based  two-dimensional (2D) systems was analyzed through magnetoconductance.~\cite{Meijer2004Nov} 
In the 2D electron gas formed at LaAlO$_3$/SrTiO$_3$ interfaces, a combination of spin-orbit coupling and scattering by finite-range impurities gives rise to a single particle mechanism of positive magnetoconductance in response to in-plane magnetic fields and at temperatures up to the 20K range~\cite{Diez2015}. Furthermore, the negative longitudinal magnetoresistance of Weyl semimetals~\cite{Son2013} is connected to the chiral anomaly of Weyl fermions. It can reach values up to $40\%$ and exhibits a strong angular dependence~\cite{Huang2015,Arnold2016}.

In nanostructures, geometrical effects owing to atomic structures,~\cite{Odom1998Jan,Odom2000Apr} strain- and defect-engineering~\cite{Santra2024May} or layer stacking~\cite{James2023Apr} can be the platform for magnetoresistive phenomena as well
~\cite{Gentile2022Sep}.
For instance, the ballistic magnetoconductance calculated in carbon nanotube reveals a step-like structure as a function of magnetic flux.~\cite{Lin1995Mar} 
In topological insulators (TI) nanowires, the $\pi$ Berry phase due to the spin-momentum locking of the surface states leaves its hallmark on the electronic band structure and provides a gap in the energy spectrum~\cite{Zhang2009}. When threaded by a half magnetic flux quantum, the surface state gap effectively vanishes~\cite{Zhang2010}, 
 thereby implying a positive magnetoconductance~\cite{Cho2015,Du2016} and Aharonov-Bohm oscillations. 
 Additionally, magnetotransport has been theoretically studied in shaped TI nanowires, such as nanocones and dumbbells~\cite{Graf2020Oct,Kozlovsky2020Mar}, where the surface electrons experience an out-of-plane component of the coaxial magnetic field. This variation in cross-sectional area leads to unconventional magnetic transport properties.
Furthermore, geometrical effects have been also shown to lead to dipolar distributions of Berry curvature~\cite{Ho2021Feb,Sod2015Nov,Ma2019Jan}
and consequently to the observation of a non-linear Hall effect in the presence of time-reversal symmetry~\cite{Ho2021Feb}.

In this study, we focus on the magnetotransport properties of carbon nanoscrolls with a turn number of two or fewer ~\cite{Xie2009Jul, Wang_2024, Shi_2010,Perim_2014}. 
This compact nanoarchitecture can be synthesized by rolled-up technology and can be seen as radial superlattices due to their spiral cross section. This results in a very peculiar bandstructure and transport behavior different from conventional flat 
nanostructures~\cite{Chang2017May,Cui2018Apr}. Both blue phosphorous~\cite{Wang2020Oct} 
and black phosphorus nanoscrolls~\cite{Wang2018Dec} 
are characterized by high carrier mobility. Importantly, there has been growing attention on aluminum- and lithium-based batteries that make use of carbon-based radial-superlattice cathodes~\cite{Yen2016Dec,Li2019Jul,Pomerantseva2019Nov,Liu2018Aug,Geng2020Mar}.


The main findings of our study are summarized in~\textbf{Fig. 1}. 
 In a two-turn 
 carbon nanoscroll (CNS) with zigzag edges (see~\textbf{Fig. 1(a)}), the ballistic conductance is tripled when the nanostructure is threaded by a half-integer magnetic flux quantum $\phi_0$/2  (see~\textbf{Fig. 1(b)}). This translates in a positive magnetoconductance coefficient (PMC) that reaches 200$\%$. Remarkably, at low carrier densities the ballistic conductance of a carbon nanoscroll is only weakly affected by disorder (see~\textbf{Fig. 1(c)}). This is in sharp contrast to a graphene zigzag ribbon that displays a zero-conductance dip.~ \cite{Li2008Feb}

\begin{figure}[tb]
\includegraphics[scale=.5]{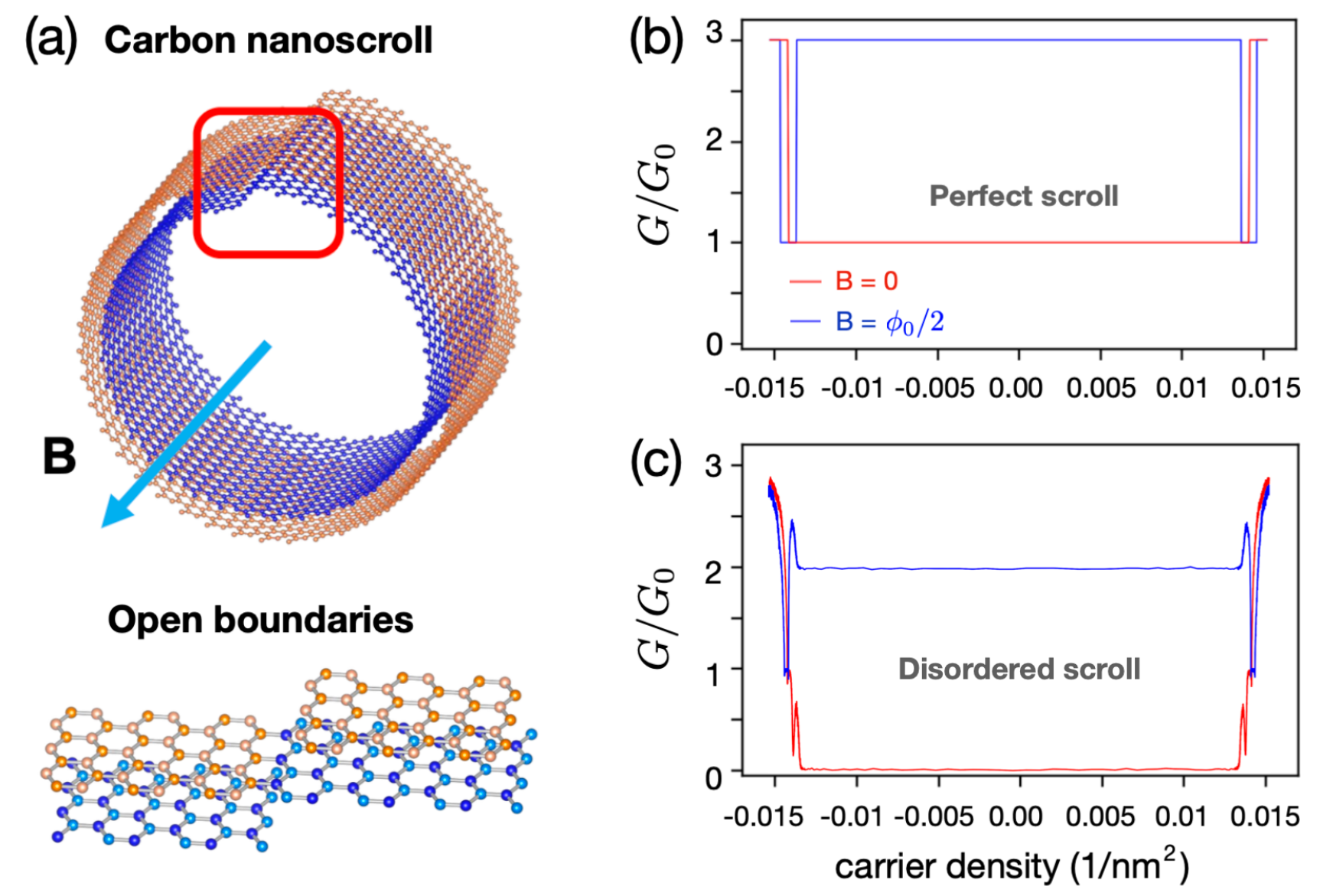}
\caption{The conductances of two-turn CNSs with and without applied magnetic fields. (a) The way of applying magnetic fields and the boundary conditions of interface. 
(b) The conductances for the case of without disorder. 
(c) The conductances for the case of with disorder. In (b) and (c), the red line
indicates the applied magnetic field $B=0$ Tesla, whereas the blue line is for $B\approx 10$ Tesla (10.3949 Tesla in the numerical calculation).}
\label{Fig1:conductance_magnetic}
\end{figure}

\section{Methods}

In order to analyze ballistic transport in carbon nanoscrolls (CNS), we employ both a continuum ${\bf k \cdot p}$ model~\cite{Zhong2022Jan}  and a tight-binding model, with which we perform numerical calculations using the Kwant package~\cite{Groth2014Jun}. In the following, we consider a two-turn CNS that can be mapped to bilayer graphene (see~\textbf{Fig. 1(a)}) with mixed boundary conditions~\cite{Zhong2022Jan}. The corresponding four-band continuum model takes into account the sublattice and layer degrees of freedom and can be written in the $A1, B1, A2, B2$ basis~\cite{McCann2013}. The resulting energy dispersion can be obtained from the relation
$\hbar vk_{\pm}=\sqrt{\varepsilon^{2}\pm \gamma_{1}\varepsilon-\hbar^{2}v^{2}k_{z}^{2}}$, 
where $k_{\pm}$ and $k_z$ are the momenta in the tangential and axial direction of the CNS respectively. In the equation above we introduce the velocity $v=\sqrt{3}a\gamma_{0}/2\hbar$  with lattice constant $a$ (see Section S1 
in Supporting Information). 

We construct a tight-binding model for a carbon nanoscroll by rolling a zigzag graphene nanoribbon perpendicular to its edges, restricting to AB-stacked (Bernal stacked) structures. The model accounts for nearest-neighbor and interlayer hoppings. 
The unit cell consists of pairs of A-B carbon atoms from the nanoribbon, represented as $\{A_1, B_1; A_2, B_2;$ $\ldots; A_{m}, B_{m}\}$, as shown in~\textbf{Fig.~2(a)}, where $m$ is the number of pairs of A-B carbon atoms in the unit cell. 
Along the zigzag boundaries, the carbon atoms of different layers are also aligned according to the AB-stacking configuration. 
To form a two-turn CNS with an AB-stacked structure, an example with 7 pairs of A-B carbon atoms ($m = 7$) is presented in ~\textbf{Figs.~2(b) and (c)}~\cite{Cong2011Nov,Yan2011Mar}. In this structure, half of the atoms sit above the centers of the hexagons, while the others are directly above the atoms of the inner layer.

\begin{figure}[h!]
	\includegraphics[scale=0.5]{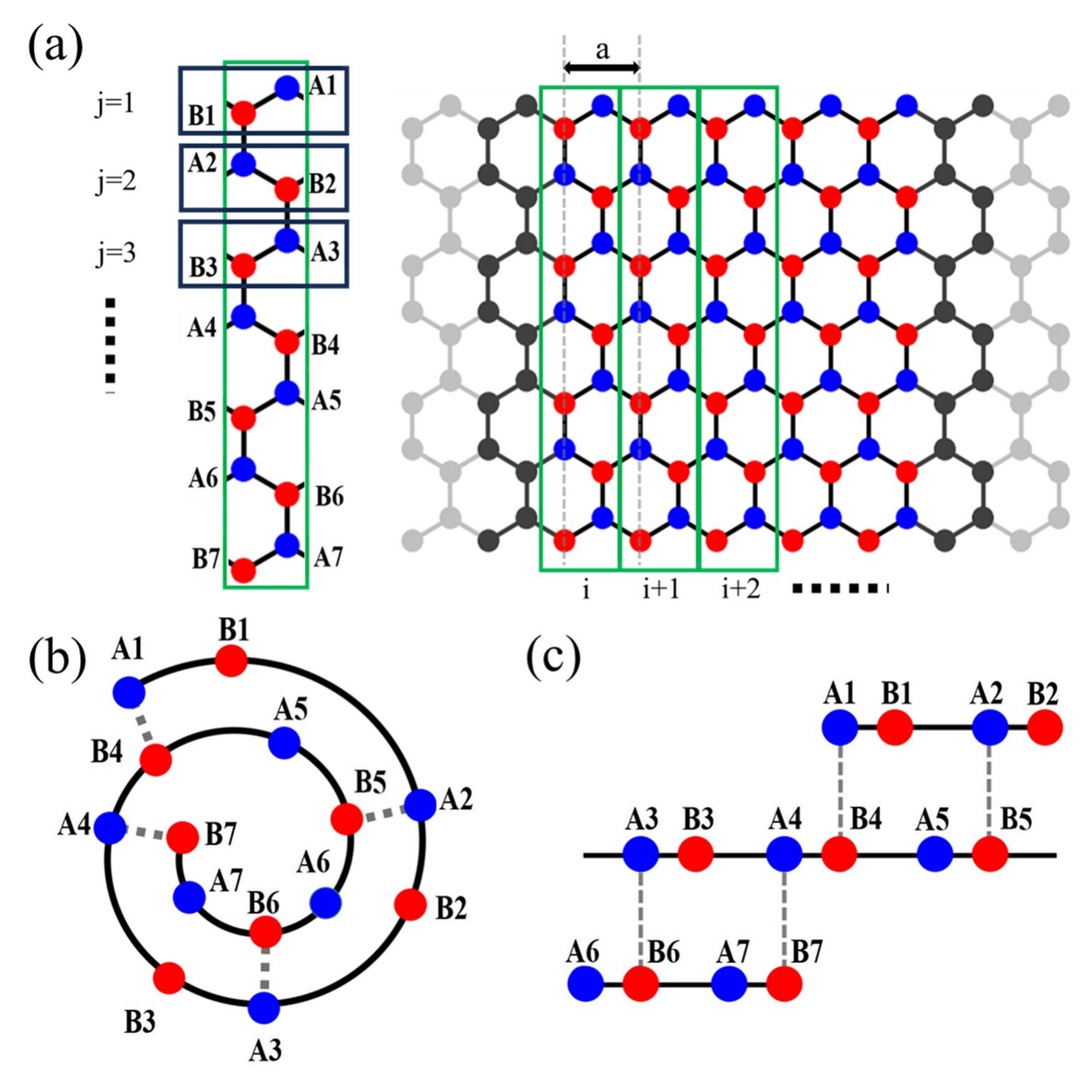}
\caption{(a) The unit cell, as denoted by green rectangular, of CNS is defined along the axis perpendicular to the core axis and has open boundaries. The length of the unit cell is determined by the arclength of the nanoscroll. 
	(b) The cross section of a CNS featuring two turns. (c) The flattened interlayer structure of the CNS shows an AB-stacking arrangement. The dashed lines in (b) and (c) denote the interlayer coupling $\gamma_1$.} 
\label{CNS_structure_unitcell}
\end{figure}

For modeling the two-turn CNS, we fix the intralayer coupling strength between A and B sites at $\gamma_{0} = 3.16$ eV,
and the interlayer coupling strength between site A2 (A site in the 2nd turn) and site B1 (B site in the 1st turn) with  $\gamma_{1}=0.381$ eV, corresponding to an interlayer distance of  $3.35$ \AA~ in the AB-stacked bilayer graphene.~\cite{Kuzmenko2009Oct,McCann2013,Gava2009Apr} The lattice
constant $a$ is 2.4595 \AA ~\cite{Saito1998Jul} with carbon-carbon bond length 1.42 \AA ~for graphene.
For the system length scale, we set the total arc length to $X = 100$ nm, which contains 934 atoms, for both the two-turn nanoscroll and the  M\"{o}bius  tube in the Kwant simulation.~\cite{Groth_2014} This corresponds to a perimeter of $L = 50$ nm and a radius of $7.99$ nm for a single turn. Furthermore, the length along the core axis is $300$ nm.

We define the positive magnetoconductance coefficient (PMC) 
as
\begin{equation}
\text{PMC}=\frac{G(\phi)-G(0)}{G(0)},
\end{equation}
where $G(\phi)$ indicates the conductance with magnetic flux $\phi$.
 The two-terminal conductance in the ballistic regime is given    
by the Landauer formula~\cite{Landauer1957Jul,Bagwell1989Jul,Chang2014Nov,Zhong2022Jan}, 
\begin{equation}
G(E_F,T)=\int_{-\infty}^{\infty}G(E,0)\left.\frac{\partial f}{\partial E}\right|_{E_F} dE,
\end{equation}
where $f$ is the Fermi-Dirac distribution function, and $E_F$ is the Fermi energy. 
The zero temperature perfectly ballistic conductance of our one-dimensional nanostructure is proportional to the number of modes ($N_s$) and given by 
$G(E,0)=2 e^2 N_s /h $. 
We neglect the mild spin-orbit coupling of graphene.

To account for the effect of disorder, we include a random on-site potential that is  Gaussian distributed~\cite{Zhong2022Jan,Bardarson2010Oct} (see Section S2 
in Supporting Information). We consider two characteristic disorder strengths of  $0.1$ eV and $0.5$ eV respectively, both comparable to the intralayer hopping amplitude. 
We examined the convergence of the averaged conductance and found that 200 configurations already achieve a small fluctuation of $5\%$. Therefore, we use 200 random disorder configurations for the results presented in the article, unless otherwise stated. 
More details on disorder convergence tests and, in addition, the calculations 
for the localization length, proportional to the mean free path in a (quasi-) one-dimensional system,~\cite{ECONOMOU2005444} are provided in Section S2
of the Supporting Information.

\section{Results and discussion}


\begin{figure}[h!]
\includegraphics[scale=.7]{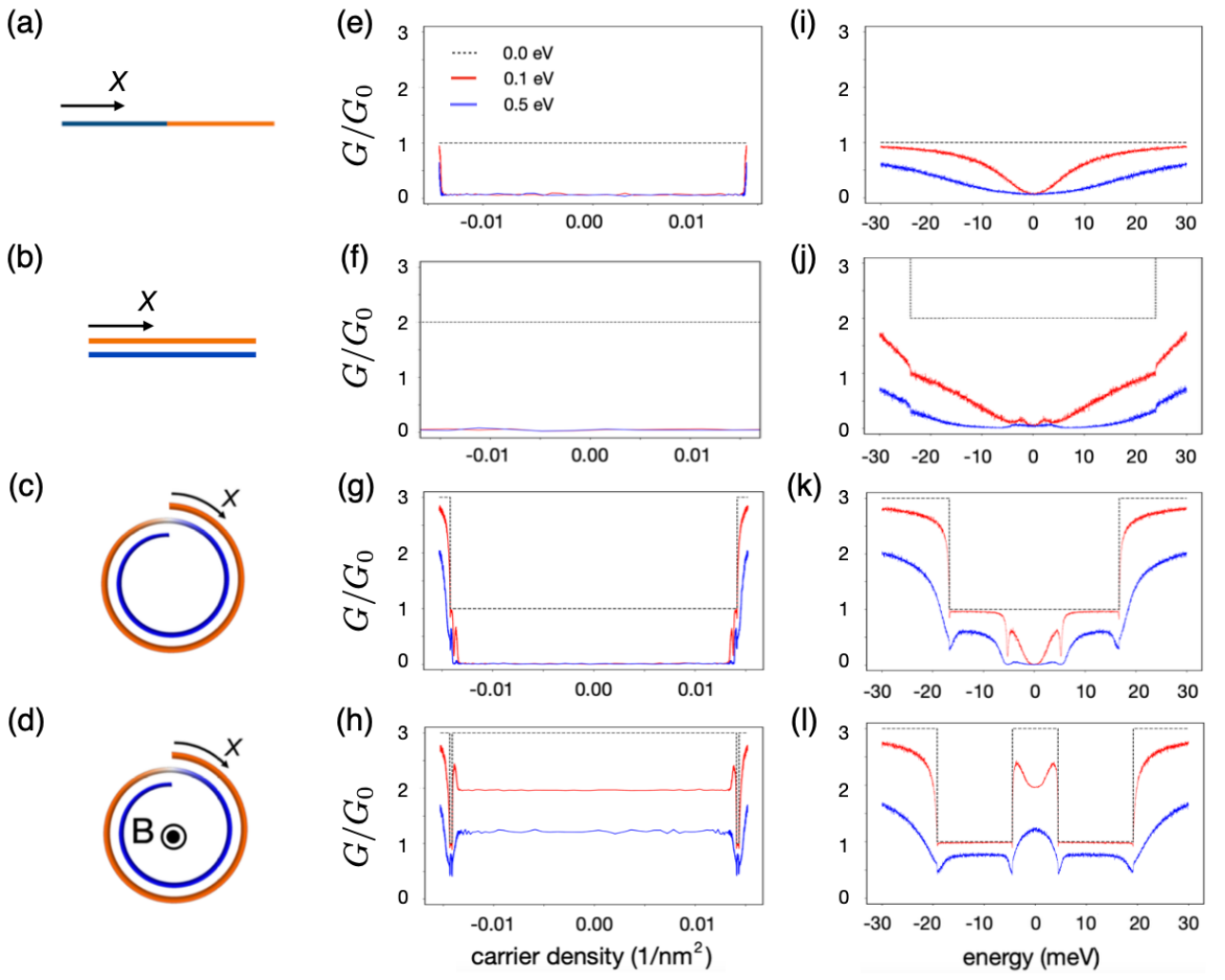}
\caption{ The conductances of  (a) monolayer ribbon, (b) bilayer ribbon, (c) two-turn CNS and (d) two-turn CNS with applied magnetic flux for different disorder strengths. 
(e) The conductances of a monolayer ribbon, i.e. 
$\gamma_1=0$ eV. (f) The conductances of a AB-stacked bilayer ribbon with interlayer coupling strength $\gamma_1=0.381$ eV. (g) The conductances of a two-turn CNSs without applied magnetic fields. (h) The conductances of two-turn CNSs with applied magnetic fields $B\approx 10$ Tesla (10.3949 Tesla in the numerical calculation). As the counterparts of (e) to (h), (i) to (l) shows the conductance with the x-axis resenting energies. The gray dashed line denotes the conductance for the perfect lattice. The red and blue line 
indicates the result of a system with applied disorder $0.1$ eV and $0.5$ eV, respectively.  } 
\label{Fig3:robust_conductance}
\end{figure}

To get a comprehensive understanding of the transport properties of a two-turn CNS threaded by a magnetic flux, we first study the two-terminal conductance of a monolayer graphene nanoribbon with zigzag edges and ribbon width equal to the total arclength of our two-turn nanoscroll (see~\textbf{Fig. \ref{Fig3:robust_conductance}(a)}). 
Based on order-of-magnitude estimation, we expect that the Zeeman effect and spin-orbit coupling have a negligible impact on the large PMC~\cite{Banszerus2020May,note_Zeeman}.
~\textbf{Figures \ref{Fig3:robust_conductance}(e) and (i)} show that in the low carrier density regime (electron or hole carrier density lower than 0.015 nm$^{-2}$) , and thus close to the charge neutrality point (Fermi energy $|E| < 30$ meV), the ballistic conductance is simply given by $G_0$. 
Disorder leads to a zero-conductance dips close to the charge neutrality point (see red and blue lines in~\textbf{Figs. \ref{Fig3:robust_conductance}(e) and (i)}). 
Very similar features are encountered when considering either a bilayer graphene ribbon  (see~\textbf{Figs. \ref{Fig3:robust_conductance}(f) and (j)}), a two-turn CNS in the absence of externally applied fields (see~\textbf{Figs. \ref{Fig3:robust_conductance}(g) and (k)}), or a double-walled carbon nanotube (see Section S3 in Supporting Information). 

For a CNS threaded by a half-integer magnetic flux quantum (see~\textbf{Fig. \ref{Fig3:robust_conductance}(d)}), the situation is completely different. As shown in ~\textbf{Figs. \ref{Fig3:robust_conductance}(h) and (l)}, the ballistic conductance at charge neutrality is tripled, comparing to the results of monolayer nanoribbon shown in~\textbf{Figs. \ref{Fig3:robust_conductance}(e) and (i)}. 
Furthermore, adding disorder does not lead to any zero-conductance dip even for disorder strength of about $0.5$ eV, and thus larger than the interlayer hopping amplitude (see the blue line in~\textbf{Figs. \ref{Fig3:robust_conductance} (h) and (l)}). We thus find that a CNS is characterized by a PMC that reaches 200\% near the charge neutrality point.  
Moreover, the localization length along the core axis of a two-turn CNS with a magnetic flux exceeds one micrometer, as detailed in Section S2 of the Supporting Information. This indicates that PMC can be realized in nanoscroll systems with length scales ranging from nanometers to micrometers \cite{Xie2009Jul, Perim_2014}.

We note that the additional phase of the nanoscorll states, determined by the applied magnetic flux, is given by $\phi=\pi(\frac{L}{2\pi} )^2 B$, where $B$ is the magnetic field strength and $L$ is the one-turn length of the nanoscroll. For our proposed magnetotransport to occur at $\phi=\pi/2=\pi (L/2\pi)^2B_c$, the required magnetic field strength $B_c$ can be reduced by a factor of $N^2$ times by increasing the nanoscroll's arclength by a factor of $N$. For a two-turn nanoscroll with an arclength eof $150$ nm, for example, the required field strength, achieving the results shown in~\textbf{Figs. \ref{Fig3:robust_conductance}(h) and (l)}, can be reduced to approximately 4.6 Tesla (see Section S3 in Supporting Information).
Additionally, the energy and conductance under various applied magnetic fields are presented in Section S4 of the Supporting Information.

\begin{figure}[h!]
\includegraphics[scale=0.5]{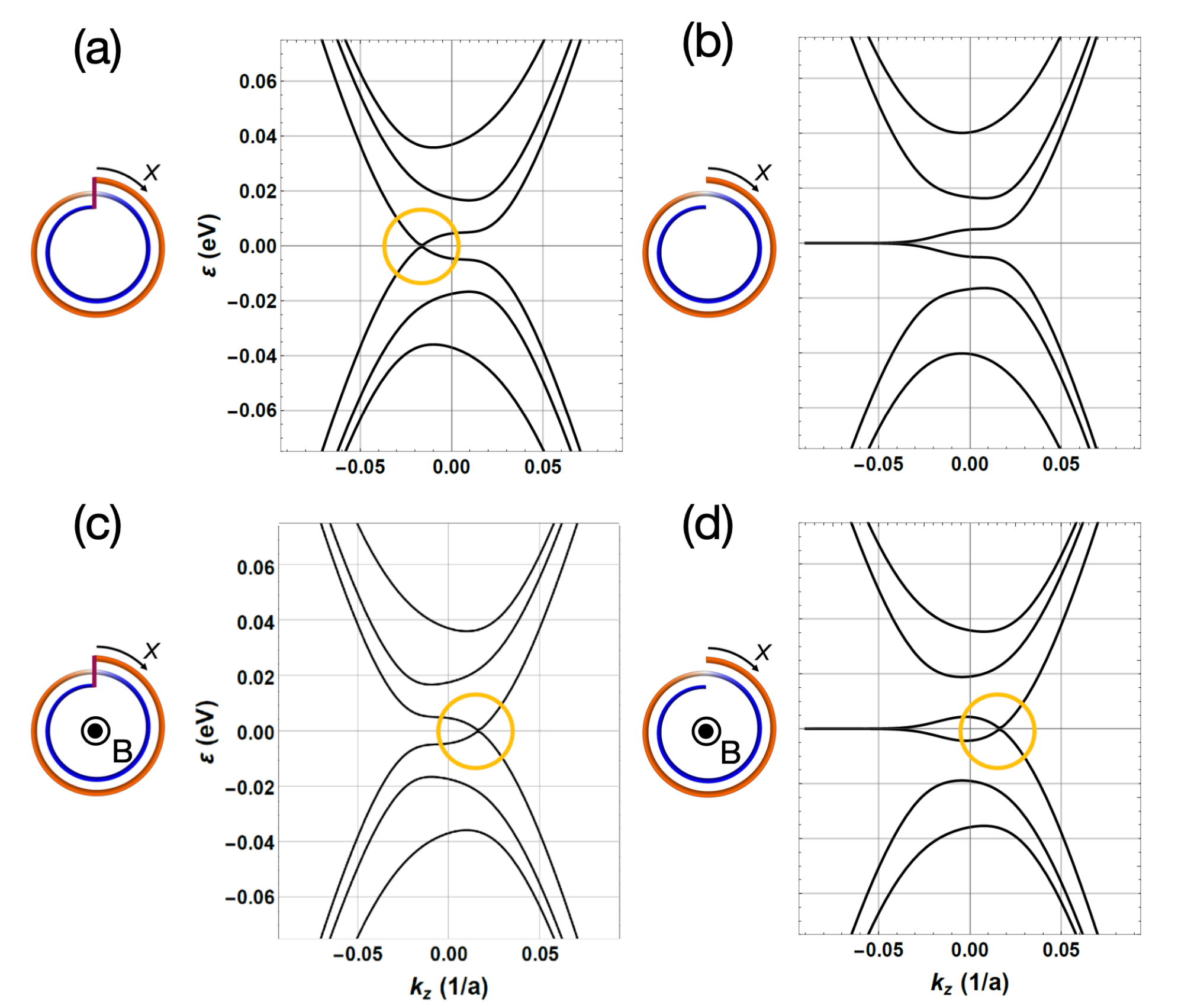}
\caption{Energy bands of the two-turn CNSs and M\"{o}bius tube for the K point: (a) M\"{o}bius tube and (b) two-turn CNSs which are without applied magnetic fields, whereas (c) M\"{o}bius tube and (d) two-turn CNSs are with applied magnetic fields $B\approx 10$ Tesla (10.3949 Tesla in the numerical calculation)}
\label{Fig4:Mobius_Nanoscrolls_bands}
\end{figure}

The conductance tripling in CNS threaded by a half-integer magnetic flux quantum can be understood by considering the electronic characteristic of CNSs. We start by considering a 
M\"{o}bius-like geometry in which the open endpoints of the CNS are closed (see~\textbf{Fig. \ref{Fig4:Mobius_Nanoscrolls_bands}(a)}). Close to the ${\bf K}$ ($K^{\prime}$) valley we observe the appearance of a two-fold degenerate zero-energy modes. This zero energy modes disappear when opening the boundary conditions as in an actual CNS (see~\textbf{Fig. \ref{Fig4:Mobius_Nanoscrolls_bands}(b)}). Instead, we observe the appearance of the characteristic zero-energy edge modes of zigzag terminated graphene. With a half-integer magnetic flux quantum 
the energy spectrum for a M\"{o}bius-like  CNS does not qualitatively change -- we only observe a shift in the axial momentum of the doubly degenerate zero energy modes (see~\textbf{Fig. \ref{Fig4:Mobius_Nanoscrolls_bands}(c)}). The case of a CNS with open boundary conditions threaded by a magnetic flux retains the zero-energy zigzag edge states found in the absence of magnetic fields. 
However, we concomitantly find the emergence of the zero energy doublet found with closed boundary conditions (see~\textbf{Fig. \ref{Fig4:Mobius_Nanoscrolls_bands}(d)}). 
It is the appearance of these additional modes that lead to the tripling of the ballistic conductance in the vicinity of charge neutrality point (the region of low carrier densities).

\begin{figure}[t]
\centering
\includegraphics[width=1.0\linewidth]{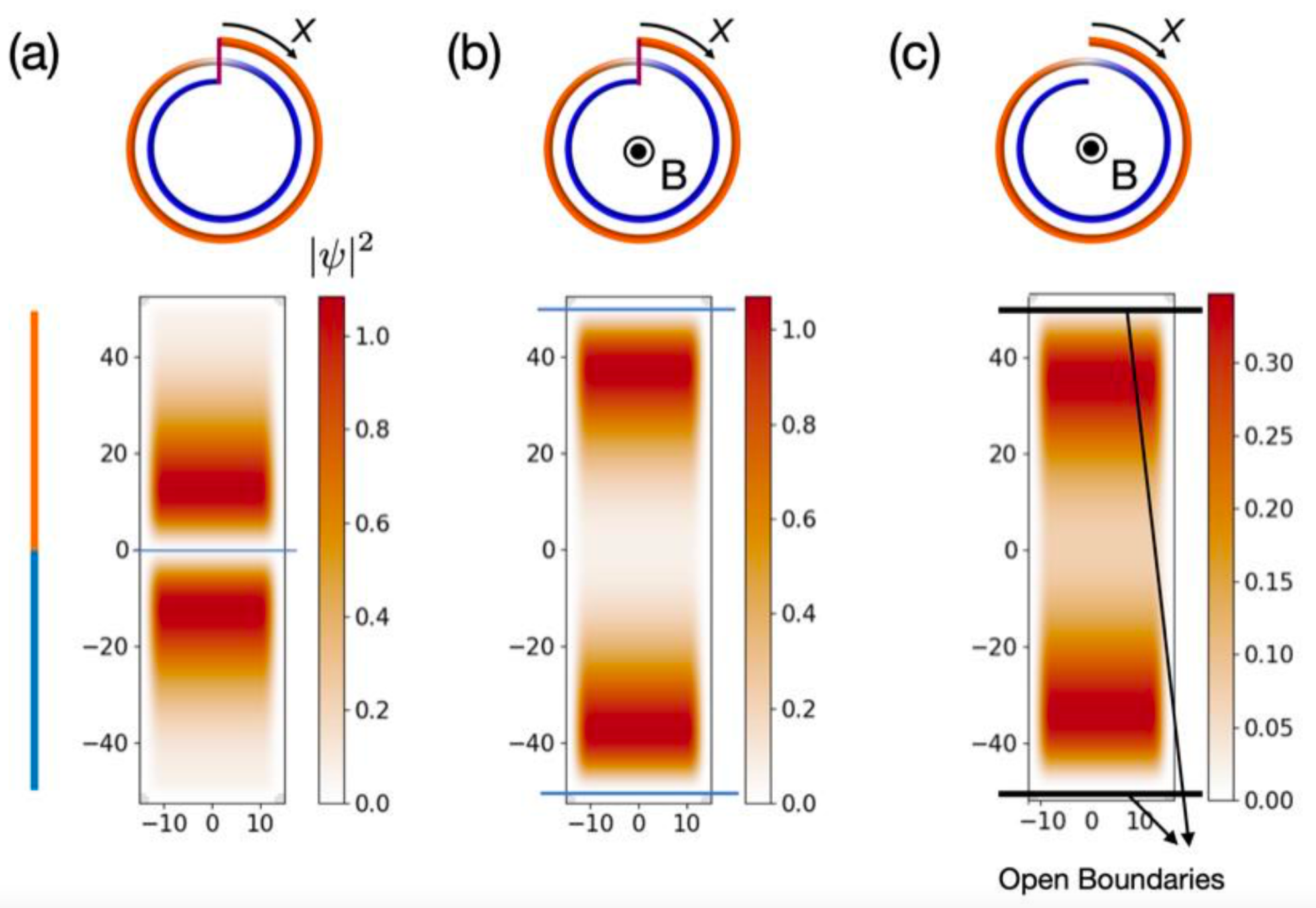}
\caption{The distributions of charge densities of the doubly degenerate zero energy states in: (a) M\"{o}bius tube, (b) M\"{o}bius tube with applied magnetic flux, and (c) CNS with applied magnetic field. The results are obtained by calculating tight-binding models of the M\"{o}bius tube and the CNS. }
\label{charge_distribution}
\end{figure}

To further demonstrate that the doubly degenerate states at zero energy in the CNS with a magnetic flux are inherited from the nontrivial interfacial states in the M\"{o}bius-like CNS, we have estimated the charge density distributions of the zero-energy states in the M\"{o}bius-like CNS, the M\"{o}bius-like CNS with an applied magnetic flux, and the CNS with the same magnetic flux. The results, shown in Fig. \ref{charge_distribution}, confirm this connection
We would like to emphasize that pioneering studies~\cite{Zhang2013Jun,Ju2015Apr,Vaezi2013Jun,Geisenhof2022Jul} have shown that the AB-BA interface in bilayer graphene induces a topological feature in $k$-space, resulting in 1D interfacial topological valley states. Our findings in Figs. \ref{Fig4:Mobius_Nanoscrolls_bands} and \ref{charge_distribution}  demonstrate that the nontrivial interfacial state is sustained not only in a  M\"{o}bius tube with the same interface but also in a carbon nanoscroll under an applied magnetic field.

\section{Conclusion}
To sum up, we have theoretically demonstrated that
 radial superlattices, especially in AB- and BA stacked domain wall featuring two-turn CNS with magnetic flux,
 display a giant PMC. 
We have found that the PMC of a two-turn CNS is giant and up to more than two times of that of the ordinary graphene nanoribbon. With simulations of disordered systems, we have found that its conductance is less prone to disorder and PMC even increases, in contrast to the disordered TI nanowire that PMC decreases remarkably~\cite{Zhang2009,Zhang2010}. 

To interpret this novel result, we developed a model of the M\"{o}bius tube with an AB-BA bilayer interface and compared its band structures and quantum states with and without magnetic flux.
The proposed PMC stems from nontrivial interfacial magnetic states, enabling it to persist not only under on-site disorder but also in systems with moderate lattice misalignment (see Section S5 in the Supporting Information) or an imperfect turn number in the nanoscroll (see Section S6 in the Supporting Information).
It is expected that the insights and effects we have unveiled in our work will be observed in the experimental field.

 \subsection*{Supporting Information:} Example: 1H NMR spectra for all compounds” or “Additional experimental details, materials, and methods, including photographs of experimental setup
 
 \section*{Author Contributions} 
 Y.-J. Z. and J.-C. Li contributed equally to this work.
  
 \section*{Acknowledgement}
We acknowledge the financial support by the National Science and Technology Council (Grant numbers 112-2112-M-006-026-, 112-2112-M-004-007 and 112-2112-M-006-015-MY2 ) and National Center for High-performance Computing for providing computational and storage resources. C.H.C. thanks support from the Yushan Young Scholar Program under the Ministry of Education in Taiwan. C.O. acknowledges support from the MAECI project “ULTRAQMAT”.  This work was supported in part by the Higher Education Sprout Project, Ministry of Education to the Headquarters of University Advancement at the National Cheng Kung University (NCKU). 

\providecommand{\latin}[1]{#1}
\makeatletter
\providecommand{\doi}
  {\begingroup\let\do\@makeother\dospecials
  \catcode`\{=1 \catcode`\}=2 \doi@aux}
\providecommand{\doi@aux}[1]{\endgroup\texttt{#1}}
\makeatother
\providecommand*\mcitethebibliography{\thebibliography}
\csname @ifundefined\endcsname{endmcitethebibliography}
  {\let\endmcitethebibliography\endthebibliography}{}

\end{document}